\documentclass[12pt,preprint]{aastex}

\slugcomment {}

\shorttitle{Identifying Nearby, Young, Late-type Stars by Means of Their Circumstellar Disks}
\shortauthors{Schneider et al.}

\begin{document}

\title{Identifying Nearby, Young, Late-type Stars by Means of Their Circumstellar Disks}

\author{Adam Schneider}
\affil{Department of Physics and Astronomy, University of Georgia,
    Athens, GA 30602}
\email{aschneid@physast.uga.edu}

\author{Inseok Song}
\affil{Department of Physics and Astronomy, University of Georgia,
    Athens, GA 30602}
\email{song@physast.uga.edu}

\author{Carl Melis}
\affil{Center for Astrophysics and Space Sciences, University of California,
    San Diego, CA 92093}
\email{cmelis@ucsd.edu}

\author{B. Zuckerman}
\affil{Department of Physics and Astronomy, University of California,
    Los Angeles, CA, 90095}
\email{ben@astro.ucla.edu}

\and 

\author{Mike Bessell}
\affil{Research School of Astronomy and Astrophysics, The Australian National University,
    Weston Creek, ACT 2611, Australia}
\email{bessell@mso.anu.edu.au}

\begin{abstract}
It has recently been shown that a significant fraction of late-type members of nearby, very 
young associations (age $\lesssim$10 Myr) display excess emission at mid-IR wavelengths indicative 
of dusty circumstellar disks. We demonstrate that the detection of mid-IR excess emission can be  
utilized to identify new nearby, young, late-type stars including two 
definite new members (``TWA 33" and ``TWA 34'') of the TW Hydrae Association.  Both new TWA 
members display mid-IR excess emission in the Wide-field Infrared Survey Explorer (WISE) catalog and 
they show proper motion and youthful spectroscopic characteristics -- namely 
H$\alpha$ emission, strong lithium absorption, and low surface gravity features consistent with known 
TWA members.  We also detect mid-IR excess -- the first unambiguous evidence of a dusty circumstellar
disk -- around a previously identified UV-bright, young, accreting star (2M1337) that is a likely member of the 
Lower-Centaurus Crux region of the Scorpius Centaurus Complex.   
\end{abstract}


\section{Introduction}
Low-mass stars are the most abundant stellar constituent in our Galaxy and are likely the typical planet 
hosts, thanks to recent radial velocity and microlensing results (e.g., \citealt{gau08}).  The 
diverse zoo of planetary systems discovered to date implies that conditions in circumstellar disks are 
intimately linked to the final planetary system architectures. In turn, circumstellar environments 
appear to be controlled by the mass of the central star they orbit (Pascucci et al. 2009 and references 
therein).  Thus, it is of great interest to understand in detail the evolution of circumstellar material
surrounding low-mass stars. 

In their analysis of mid-IR excess fractions for young, nearby associations, \cite{sch12} show that a 
significant number of M-type stars display mid-IR excess emission in stellar groups younger than $\sim$10 
Myr.  \cite{riaz12} found a similar result in their analysis of primordial disk fractions for young clusters, 
namely that disks around later-type stars remain in the primordial stage for a longer period of time than 
disks around stars of earlier spectral types.  Based on the WISE channel 4 at 22 $\mu$m, for M-type members 
with spectral types between M0 and M6, \cite{sch12} derive updated excess fractions 
of $45^{+15}_{-13}$\% for the $\sim$5-8 Myr \citep{luh04} $\eta$ Cha cluster, and $21^{+12}_{-6}$\% for members 
of the TW Hydrae Association (``TWA" -- age $\sim$8 Myr; \citealt{zuck04}).  For the $\sim$12 Myr old Beta 
Pictoris Moving Group (BPMG -- age $\sim$12 Myr; \citealt{zuck04}), no evidence was found for protoplanetary (primordial or transitional) 
disks.  Of 20 M-type stars in the BPMG in this spectral range, only one case of marginal 22 $\mu$m excess was recovered, 
coming from a well-known debris disk bearing member AU Mic.  This implies that M-type stars with spectral types between M0 and
M6 exhibiting mid-IR excess are very likely young (age $\lesssim$ 10 Myr).  

One can utilize this association of mid-IR  excess and youth as a new search method for identifying nearby, young 
($<$10 Myr), late-type stars and brown dwarfs.  Low-mass stars and brown dwarfs in this age range should all 
show additional unambiguous indicators of youth, such as strong H$\alpha$ emission, lithium absorption, low-gravity spectral features, 
etc.  Therefore, the youth of any candidate young M-type object discovered by its excess emission at mid-IR wavelengths 
can be evaluated with follow-up spectroscopy.     

Although several young ($\lesssim$100 Myr), nearby ($\leq$80 pc) moving groups were identified during the 
past decade (\citealt{zuck04}, \citealt{tor08}), these include few low-mass members (spectral types later than 
$\sim$M3).  This is mainly due to the fact that 
unambiguous identification of young M-type stars is difficult if a well-measured trigonometric parallax is 
missing.  Some activity indicators (e.g. H$\alpha$ emission, X-ray emission) 
cannot readily discern young ($\lesssim$100 Myr) stars from a pool of  old ($\geq$600 Myr) field stars \citep{shk09}.  The Li 
$\lambda$6708 absorption feature, an effective age indicator for young FGK-type stars in the age range of 10-100 Myr, 
becomes increasingly sensitive to mass and age for M-type stars.  The age at which a star becomes hot 
enough to burn lithium in its core can be seen in the so-called ``lithium depletion boundary'', which can affect 
early to mid-M type stars at ages as young as the BPMG ($\sim$12 Myr).  

Some young M-type stars have been identified, though they are either co-moving companions of earlier spectral types 
or in a group of very young ($\lesssim$10 Myr) stars that are relatively confined to a small region of the sky (e.g. 
TWA and the subregions of the Scorpius Centaurus Complex).  With a well-measured trigonometric parallax, one can
find a young M-type field star (e.g. AP Col; \citealt{rie11}), however, a systematic search for young M-type stars needs to 
wait for the next generation of parallax missions, such as {\it Gaia}, {\it Pan-STARRS}, etc.  Using the fact that random 
field M-type stars with mid-IR excess are extremely rare, one can search for $\lesssim$10 Myr M-type stars in the solar 
neighborhood with WISE data.  These additional youngest M-type post T-Tauri stars can give an important clue to the mass 
function of young nearby stellar associations.

The TW Hydrae Association is one of the nearest (d$\sim$30-90pc) star forming regions.  Its proximity,
in combination with its young age make it an ideal area for the study of stellar, planetary,
and circumstellar disk evolution.  It has also been shown to be useful as a testbed for the evaluation and 
implementation of new techniques to identify young stars (e.g. UV-excess; \citealt{rod11}, \citealt{shk11}).  
For the above reasons, we have chosen to examine the TW Hydrae Association in a pilot study to find
young, M-type stars by their mid-IR excess emission.    
               
\section{Search Method}
First, we define a search area around known TWA members in decimal degress for right ascension (R.A.) 
and declination (Dec.) (Figure 1).

150.0 $<$ R.A. $<$ 205.0

-50.0 $<$ Dec. $<$ -25.0\\

We cross-match all 2MASS sources in this region with WISE using a source matching radius of 5\arcsec, 
restricting our search to those matches that are well-detected (i.e. not upper limits) in all 2MASS and WISE 
bands, and require the WISE extended source flag to be less than 2 (i.e. not 
associated with a 2MASS extended source catalog object).  Then, we make the following magnitude cut in 
2MASS J-magnitude in an attempt to exclude extragalactic sources, and we restrict various 2MASS/WISE colors to 
be those of M-type objects.  2MASS/WISE color selection criteria for M-type objects were determined by examining 
the colors of known M-type members of the $\beta$ Pic, $\eta$ Cha, and TW Hydrae associations.   

J $<$ 14.0 (mag)

0.5 $<$ J-H $<$ 0.75 (mag)

0.2 $<$ H-K$_S$ $<$ 0.5 (mag)   

0.8 $<$ J-K$_S$ $<$ 1.2 (mag)

0.75 $<$ J-W1 $<$ 2.0 (mag)\\

The resulting list is replete with spurious detections in WISE channel 4 (for example, sources with S/N $\geq$ 
5, but no obvious source visible in W4 images).  To address this issue, we make a WISE channel
4 magnitude cut in such a way as to detect an M-type star at $>$ 5$\sigma$ based on a typical uncertainty for a detectable 
faint source of $\sim$1.0 mJy.  Using the zero-point flux from \cite{jar11}, we find a source with a flux of 5.0 mJy
in WISE channel 4 corresponds to a magnitude of $\sim$8.1 mag.  Therefore, we keep as candidates all sources with a measured
WISE channel 4 magnitude less than 8.1 mag.  

As stated previously, candidates are required to have detections in WISE channel 4.  This refers to the W4 magnitude
measured with profile-fitting photometry provided in the WISE catalog.  The WISE catalog also provides a W4 magnitude
as determined via aperture photometry, which can be found in the ``long form'' version of the WISE all-sky source catalog.  
In addition to the requirement that all candidates were well-detected in the profile fitting photometry, we require candidates 
to be well-detected (i.e. not an upper limit) as determined by their magnitude 
measured via aperture photometry.  By carefully eye-checking a selection of sources, it was found that those with an 
upper limit in WISE channel 4 determined by aperture photometry, even when the profile-fitting magnitude indicates a 
good detection, are spurious detections.  By forcing PSF fitting and aperture photometric results to be similar, approximately 
half of all remaining candidates were discarded in this way.      

WISE channel 1 and WISE channel 4 magnitudes are then used to select objects with mid-IR excess.  
As seen in Figure 2 of \cite{sch12}, a typical W1-W4 color for an M-type non-excess star is between 0 and 1.  
Objects having colors residing in the following range are selected as excess candidates.  

W1-W4 $>$ 1.0 (mag)

Mid-IR excess candidate selection is displayed in Figure 2.

A total of 76 M-type mid-IR excess candidates were found in the TWA search area defined above.  These were 
then further examined by checking the corresponding WISE images of each candidate for any evidence of 
contamination, spurious detections in WISE channel 4 (S/N $\geq$ 5, but
no obvious source visible in W4 images), or an extended shape indicative of an extragalactic nature.  
After the WISE image screening, 51 candidates remained.  This list was then cross-matched with SIMBAD.  Twelve 
sources were found to be either extragalactic or pulsating variable stars.  

One object (2MASS 13373839-4736297), is a known young M-dwarf first discovered by 
\cite{rod11} in their search for young stars utilizing UV-excess.  \cite{rod11} show that this star, with an estimated 
spectral type of M3.5 and distance of 126 pc, shows youthful characteristics in its spectrum.  They measure an 
H$\alpha$ equivalent width of 13.7 \AA in emission, and a Li $\lambda$6708 equivalent width of 308 m\AA.  This object is likely a 
member of Lower-Centaurus Crux region (d$\sim$95 pc, age $\sim$10 Myr; \citealt{song12}) of the Scorpius 
Centaurus Association, based on its estimated distance and proper motion.  This is the first evidence for mid-IR 
excess for this star, which shows that our search method is likely to be useful for other areas of the sky, such as 
the sub-regions of the Scorpius-Centarus Complex.

Six confirmed TWA members were recovered (TWA 3A, 27, 28, 30A, 31, and 32).  The lone late-type (M4), mid-IR excess 
TWA member not returned with our search method was TWA 30B.  This object did not meet our selection criteria 
because its photospheric emission is heavily obscured up to at least WISE channel 2, likely due to its edge-on 
disk geometry (\citealt{loop10b}; \citealt{sch12}).  

For the remaining 32 candidates, we cross-match their positions with the PPMXL catalog of positions and proper 
motions \citep{roe10} to inspect for any candidates that have measured proper motions consistent with that of 
known TWA members.  To select the best candidates, we require that the total proper motion magnitude and 
direction be within 1$\sigma$ of the average of confirmed TWA members.  Thirty of the candidates show 
inconsistent proper motions.  The last two remaining candidates (2MASS 11393382-3040002 and 2MASS 
10284580-2830374), with proper motions consistent with known members of TWA (see Figure 1), are discussed 
in detail in the following section.

\section{Spectroscopic Follow Up \& Membership Evaluation}

To check for the expected signatures of youth for an M-type TWA member, candidates 2M1139 and
2M1028 were followed up spectroscopically with the Wide Field Spectrograph (WiFeS; \citealt{dop07}) 
on the 2.3m telescope located at the Siding Spring Observatory.  2M1139 was observed on 2012 May 30 
and 2M1028 was observed on 2012 August 3.  The spectra were obtained with the $B_{3000}$ and $R_{3000}$ 
gratings that cover the wavelength ranges 3400 to 6000 A and 5600 to 9500 A, respectively, with a resolution of 
R $\sim$ 3000. A portion of each spectrum containing the H$\alpha$ emission line and the Li $\lambda$6708 
absorption line is shown in Figure 3 (2M1139) and Figure 4 (2M1028).  2M1139 closely resembles the average 
combined spectrum of old field dwarfs GJ 299 (M4.5) and GJ 551 (M6), while 2M1028 most closely resembles
GJ 551.  A comparison of each candidate TWA member with these spectra shows a considerable 
weakening of MgH, CaH, CaOH, KI and NaI  features compared to old field dwarfs with similar TiO band 
strengths (Figures 3 \& 4).  As seen in Table 1, each candidate shows moderate H$\alpha$ emission 
and strong lithium absorption.   

\cite{law09} show that indices derived from spectroscopic features that are sensitive to a star's surface 
gravity can be used to distinguish relative age differences of nearby clusters and moving groups.  In 
particular, for stars of spectral type M3 and later, the Na I $\lambda$8183/8195 doublet index shows a distinct 
difference between field dwarfs (strong absorption), young stars ($\lesssim$12 Myr) (reduced absorption), 
and giants, in which the feature is mostly absent.  We measure a Na I index for both candidates, and, using 
their R-I colors from our flux-calibrated spectra, compare their values with the Na I index-color relations for various 
stellar groups from \cite{law09} in Figure 5.  VRI colors (in the Johnson-Cousins system) were computed 
from our flux-calibrated spectrum for each candidate, and are listed in Table 1.  We estimate the photometry 
is accurate to within $\pm$0.01-0.02 mags. As seen in the figure, the Na I index of each candidate suggests a low surface gravity, 
and an age similar to TWA.  This low-gravity feature, along with the measured equivalent widths of the H$\alpha$ 
emission and lithium absorption lines, given in Table 1, confirm the youth of both candidates.

As mentioned in \cite{sch12}, for most cases, the determining factor between membership in TWA and the 
further away, Lower-Centaurus Crux (LCC) region of the Scorpius Centaurus Complex is distance, with a 
possible boundary between the two regions occurring near 100 pc.  We estimate the distance to each candidate 
TWA members in two ways. First, we calculate a photometric distance using our measured spectral type in 
combination with an empirical isochrone for known TWA members.  We estimate a spectral type of 
M4.7 $\pm$ 0.5 for 2M1139 and M4.9 $\pm$ 0.5 for 2M1028 using the TiO5 index as described in \cite{reid95}.  
Placing these candidates on the empirical TWA isochrone, we estimate an absolute K magnitude of $\sim$6.0 mag
and $\sim$6.5 mag for 2M1139 and 2M1028, respectively.  From these absolute magnitudes, we calculate a 
photometric distance of $\sim$41 pc for 2M1139 and $\sim$50 pc for 2M1028.  Uncertainties using this method are 
typically $\sim$20\%.  

We also estimate the kinematic distance by obtaining a kinematic parallax following the method described 
in \cite{mam05}.  The mean velocity of TWA was calculated using parameters of confirmed members from 
\cite{sch12}.  The adopted TWA space velocity is (U, V, W) = (-10.8, -18.1, -4.4) km s$^{-1}$.  As with TWA 
30A and TWA 30B, which have sky positions and proper motions similar to that 
of 2M1139 (see Figure 1), we find that almost all space motion is directed toward the TWA convergent 
point (\citealt{loop10a}; \citealt{loop10b}).  From this motion, we calculate a kinematic parallax of  23.6 $\pm$ 1.8 mas, 
corresponding to a distance of 42 $\pm$ 3 pc.  For 2M1028, we calculate a kinematic parallax of  20.3 $\pm$ 1.7 mas, which corresponds to a distance of 49 $\pm$ 4 pc.  Both estimates are in excellent agreement with our calculated photometric 
distances.  These distance estimates put each candidate well on the TWA side of the $\sim$100 pc dividing line between 
TWA and the LCC.

\cite{rod11} and \cite{shk11} show that UV-excess is an effective tool for identifying young, M-type objects.  We 
searched for any evidence of UV-excess using the Galaxy Evolution Explorer (GALEX; \citealt{mar05})
and found that 2M1139 has not been covered in the most recent data release.  2M1028 was detected by GALEX, 
with a near ultraviolet (NUV) magnitude of 22.325 $\pm$ 0.397 mag.  With a 2MASS
J-K color of 0.93 mag, and a NUV-J color of 11.37 mag, we note that this object would pass the selection criteria outlined in 
\cite{rod11} for young stars candidates, namely NUV-J $\leq$ 10.20(J-K) + 2.2, indicative of a young age.       

The combination of position, proper motion, spectral signatures of youth, distance estimates, mid-IR excess, and, in the
case of 2M1028, UV-excess,  lead us to the conclusion that 2MASS 1139 and 2M1028 are indeed authentic members 
of the TW Hydrae Association.  Following the naming convention for TWA members, we designate 2M1139 TWA 33 and 
2M1028 TWA 34.  The properties of TWA 33 and TWA 34 are summarized in Table 1.  Figure 6 includes a spectral energy 
distribution (SED) that shows the mid-IR excess emission of each new member.  Also shown is the SED of a likely LCC 
member 2M1337, first identified as a young star by \cite{rod11}, the infrared excess of which was discovered in our search.     

\section{Discussion}        
We note that, at the age of TWA or younger, use of mid-IR excess as a search method is highly effective for TW Hydrae members 
with spectral types of M4 or later.  Including TWA 33 and TWA 34, there are 16 TWA members with spectral types in this 
range.  Fourteen of these were individually detected with WISE (TWA 3B and 5B are too close to their primary
members to be resolved individually).  Nine of these fourteen members show significant mid-IR excess 
emission, eight of which were found with our search method.  The discovery of TWA 33 and TWA 34 demonstrates
the effectiveness of using mid-IR excess as a tool to identify nearby, young, late-type stars.   These stars show many signatures 
of youth expected for a TWA member, such as H$\alpha$ emission, lithium absorption, and low surface gravity spectral
features.  Using two different methods, we estimate a distance of $\lesssim$ 50 pc for both stars, consistent with their TWA
membership.   We also rediscovered the young, likely LCC member 2M1337, first identified as a 
young star by \cite{rod11}, and found evidence for the presence of a circumstellar disk 
around this star.

Considering the high fraction of late-type mid-IR excess stars from TWA and the slightly younger $\eta$ 
Cha Association (45\% for M-type members - Schneider et al. 2012), we believe that mid-IR excess can 
be a  a useful tool in identifying more young, nearby, late-type stars and brown dwarfs.  One can expand 
our search for these objects to other areas of interest, such as the sub-regions of the Scorpius-Centaurus 
Association, or even the entire sky. 

An interesting conjecture is whether there is a higher fraction of mid-IR excess stars among the latest M spectral
types (M6 or later).  For the six known TWA members with spectral type later than M6, three -- TWA 27, 30A and 
32 -- were detected individually by WISE to have excess emission in channel 4.  TWA 5B, an M8 dwarf companion 
to TWA 5A, is too close to its primary to be resolved.  TWA 26 and TWA 29 have only upper limits in WISE channel 4, 
so no strong conclusions regarding mid-IR excess can be deduced from WISE for these members, though TWA 26 
shows no excess at 24 $\mu$m via Spitzer MIPS \citep{sch12}.  In summary, all WISE detectable late M-type TWA members show
mid-IR excess emission at the W4 band.  To date, no members of the $\eta$ Cha cluster have been found
with spectral types later than M6, though searches have been performed \citep{lyo06}.  The BPMG has two known
late-type members (excluding giant planet $\beta$ Pictoris b), HR 7329B (M7.5; \citealt{low00}) and 2MASS 0608-27 
(M8.5; \citealt{rice10}).  HR 7329B is too close to its host star to be resolved individually with WISE, and 2MASS 0608-27
has an upper limit in WISE channel 4.  So, at this point, only TWA has a reasonable number of latest M-type members
to check the conjecture of even more prolonged disks among the latest M-type stars.  Future missions, such as {\it Gaia}, 
which should discover more late-type members of these nearby associations, will allow us to test the hypothesis 
that primordial disks have longer lifetimes around later-spectral types in this spectral type range.

While we have shown that mid-IR excess emission can be a useful tool for identifying nearby, young, late-type
stars and brown dwarfs, caution must be taken when considering association properties, such as disk fraction and 
mass function.  Any objects found with this method will surely bias the measure of disk frequency for a particular
group of stars, so we do not update the excess fraction for TWA here.  If the hypothesis of longer disk lifetimes 
around later spectral types is true, then any objects found with this method would bias any estimate of the 
mass function as well.  While the question of whether or not there is a higher fraction of mid-IR excess for the 
latest spectral types is an interesting one, it cannot be answered with objects found with the search method 
described in this paper.

\acknowledgments

The authors would like to thank Dr. Patrick Tisserand for obtaining the spectrum of 2M1028.  This research 
was funded in part by NASA grants to the University of Georgia and UCLA.  C. M. acknowledges support
from the National Science Foundation under award No. AST-1003318.  
This research has made use of the SIMBAD database and VizieR catalog access tool, operated at CDS, 
Strasbourg, France.  This publication makes use of data products from the Two Micron All Sky Survey, 
which is a joint project of the University of Massachusetts and the Infrared Processing and 
Analysis Center/California Institute of Technology, funded by the National 
Aeronautics and Space Administration and the National Science Foundation, as well as the {\it Wide-field
Infrared Survey Explorer}, which is a joint project of the University of California, Los Angeles, and 
the Jet Propulsion Laboratory/California Institute of Technology, funded by the National 
Aeronautics and Space Administration.

\clearpage
\begin{deluxetable}{lccc}
\tablecaption{Properties}
\tablewidth{0pt}
\tablehead{
\colhead{Parameter} & \multicolumn{2}{c}{Value} & \colhead{Ref.}\\
\cline{2-3}\\
& \colhead{TWA 33 (2M1139)} & \colhead{TWA 34 (2M1028)} }
\startdata
$\alpha$ (J2000) & 11:39:33.83 & 10:28:45.80 & 1\\ 
$\delta$ (J2000) & -30:40:00.3 & -28:30:37.4 & 1\\ 
$\mu$$_{\alpha}$ & -88.7 $\pm$ 6.1 (mas/yr) & -65.5 $\pm$ 4.1 (mas/yr) & 2\\
$\mu$$_{\delta}$ & -25.9 $\pm$ 6.1 (mas/yr) & -11.1 $\pm$ 4.1 (mas/yr) & 2\\
Optical SpT & M4.7 $\pm$ 0.5 & M4.9 $\pm$ 0.5 & 3\\
Distance\tablenotemark{a} & 41 $\pm$ 8 pc & 50 $\pm$ 10 pc & 3\\
Distance\tablenotemark{b} & 42 $\pm$ 4 pc & 49 $\pm$ 4 pc & 3\\
V - R & 1.37 $\pm$ 0.03 (mag) & 1.65 $\pm$ 0.03 (mag) & 3\\
R - I & 1.80 $\pm$ 0.03 (mag) & 1.97 $\pm$ 0.03 (mag) & 3\\
J & 9.985 $\pm$ 0.021 (mag) & 10.953 $\pm$ 0.027 (mag) & 1\\
H & 9.414 $\pm$ 0.023 (mag) & 10.410 $\pm$ 0.024 (mag) & 1\\
K$_S$ & 9.118 $\pm$ 0.023 (mag) & 10.026 $\pm$ 0.022 (mag) & 1\\
W1 & 8.765 $\pm$ 0.022 (mag) & 9.592 $\pm$ 0.023 (mag) & 4\\
W2 & 8.404 $\pm$ 0.019 (mag) & 9.116 $\pm$ 0.020 (mag) & 4\\
W3 & 7.135 $\pm$ 0.017 (mag) & 8.243 $\pm$ 0.020 (mag) & 4\\
W4 & 5.518 $\pm$ 0.038 (mag) & 6.845 $\pm$ 0.075 (mag) & 4\\
Li EW & 470 m\AA& 370 m\AA& 3\\
H$\alpha$ EW & -5.8 \AA & -9.6 \AA & 3\\
Na I index & 1.14 $\pm$ 0.02 & 1.12 $\pm$ 0.02 & 3\\
NUV & & 22.325 $\pm$ 0.397 (mag) & 5\\
\enddata
\tablenotetext{a}{Photometric distance using empirical TWA isochrone.}
\tablenotetext{b}{Kinematic distance assuming TWA membership.}
\\
References: (1) 2MASS catalog \citep{cut03}; (2) PPMXL catalog \citep{roe10}; (3) This work; (4) WISE All-Sky 
Source Catalog \citep{cut12}; (5) GALEX catalog\\
 \end{deluxetable}

\begin{figure}
\plotone{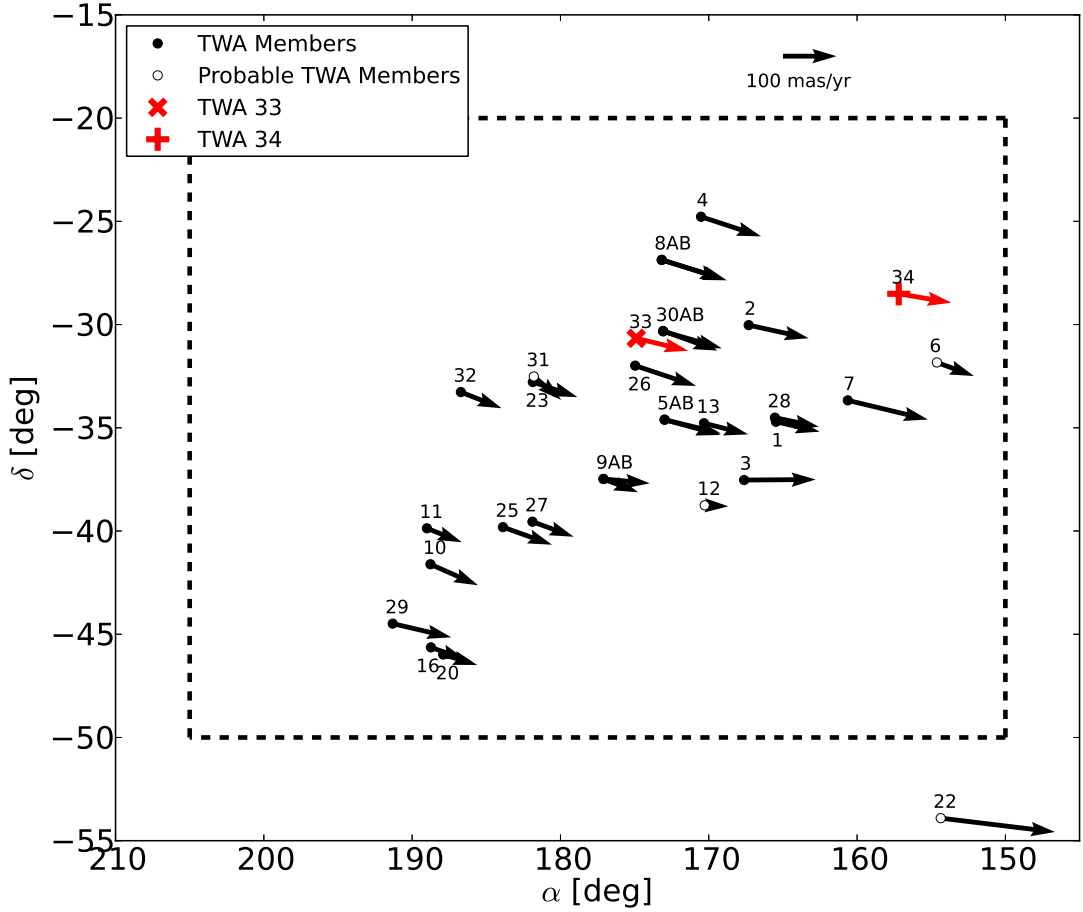}
\caption{The location of TWA 33 and TWA 34 alongside other TWA members.  Probable 
members (6,12, 22, and 31; \citealt{sch12}) are displayed as open circles.  Arrows indicate the direction and magnitude of the measured 
proper motions for each member.  The large dashed box indicates the search area defined in Section 2.}
\end{figure}
\begin{figure}
\plotone{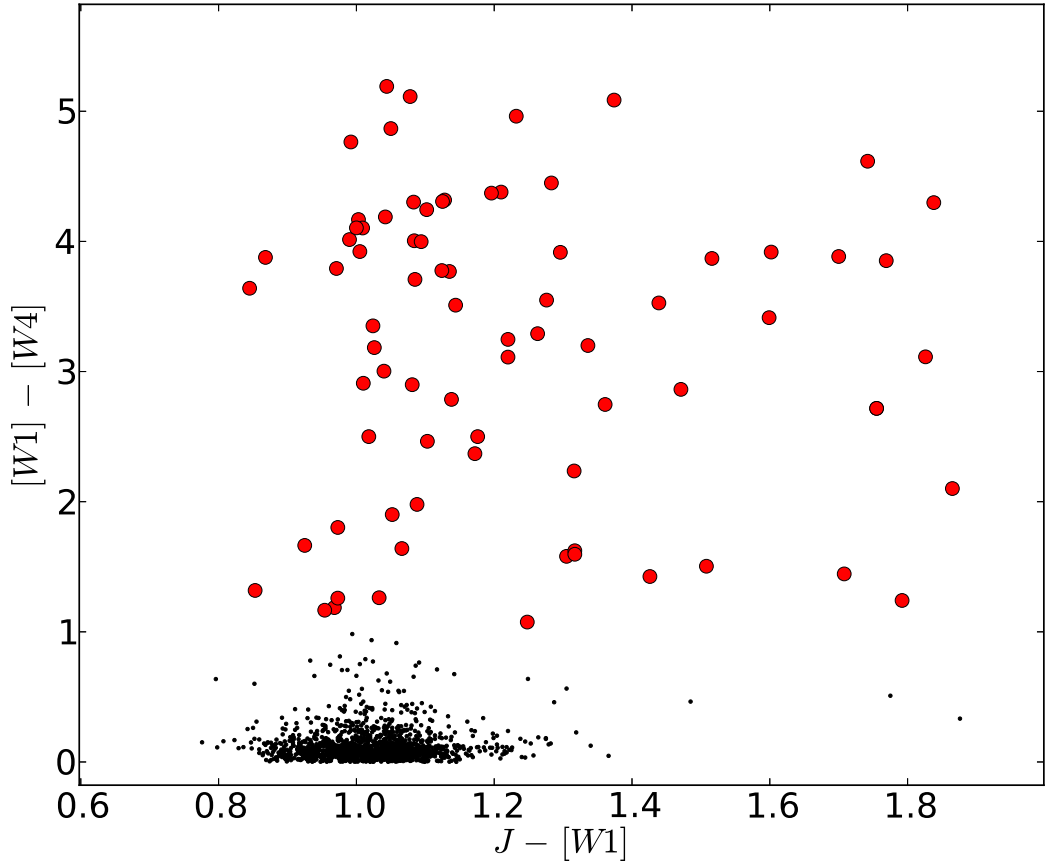}
\caption{J-W1 vs W1-W4 color-color diagram showing our excess candidate selection.  Seventy-six excess candidates are shown as large red symbols, while small black symbols represent non-excess candidates (see Section 2).}
\end{figure}
\begin{figure}
\plotone{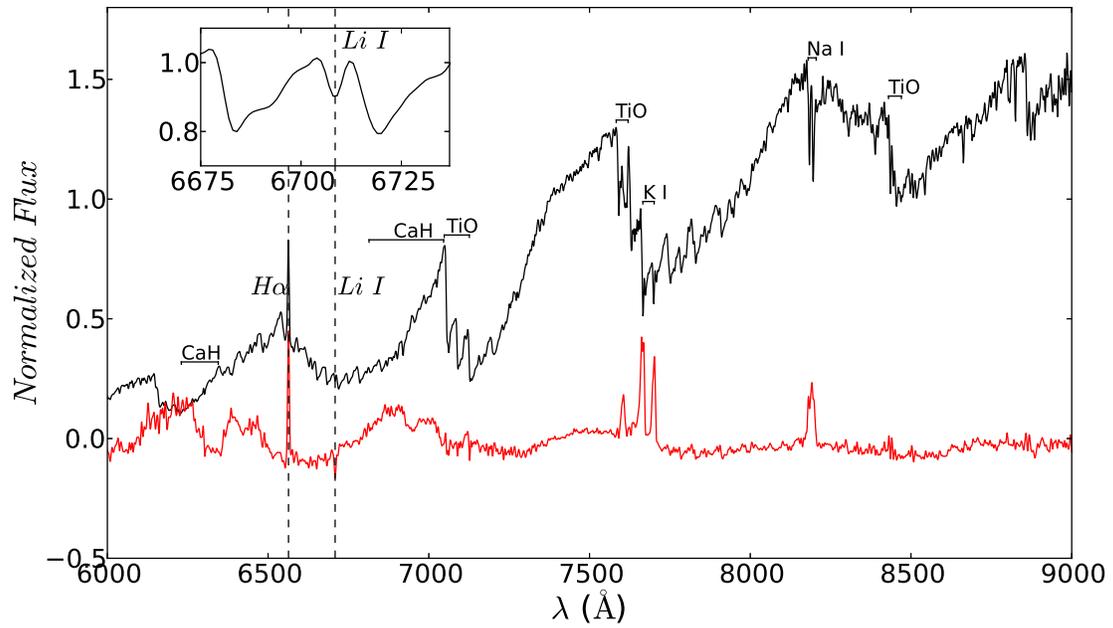}
\caption{A portion of the optical spectrum of TWA 33.  Major atomic and molecular features have been labeled.  
The inset shows its lithium absorption feature, and has been normalized to the continuum.  The red line is the
spectrum of TWA 33 divided by the combined spectra of old disk stars GJ 299 and GJ 551, highlighting 
the various gravity sensitive features.}
\end{figure}
\begin{figure}
\plotone{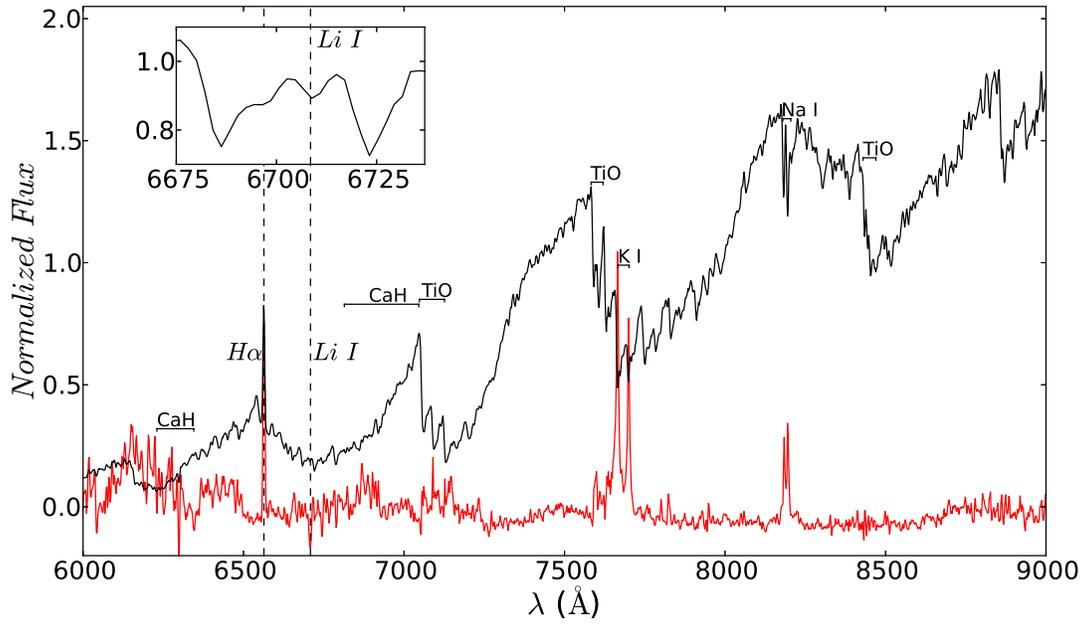}
\caption{A portion of the optical spectrum of TWA 34, displayed in the same manner as Figure 3.  The red line
is the spectrum of TWA 34 divided by the spectrum of GJ 551.}
\end{figure}
\begin{figure}
\plotone{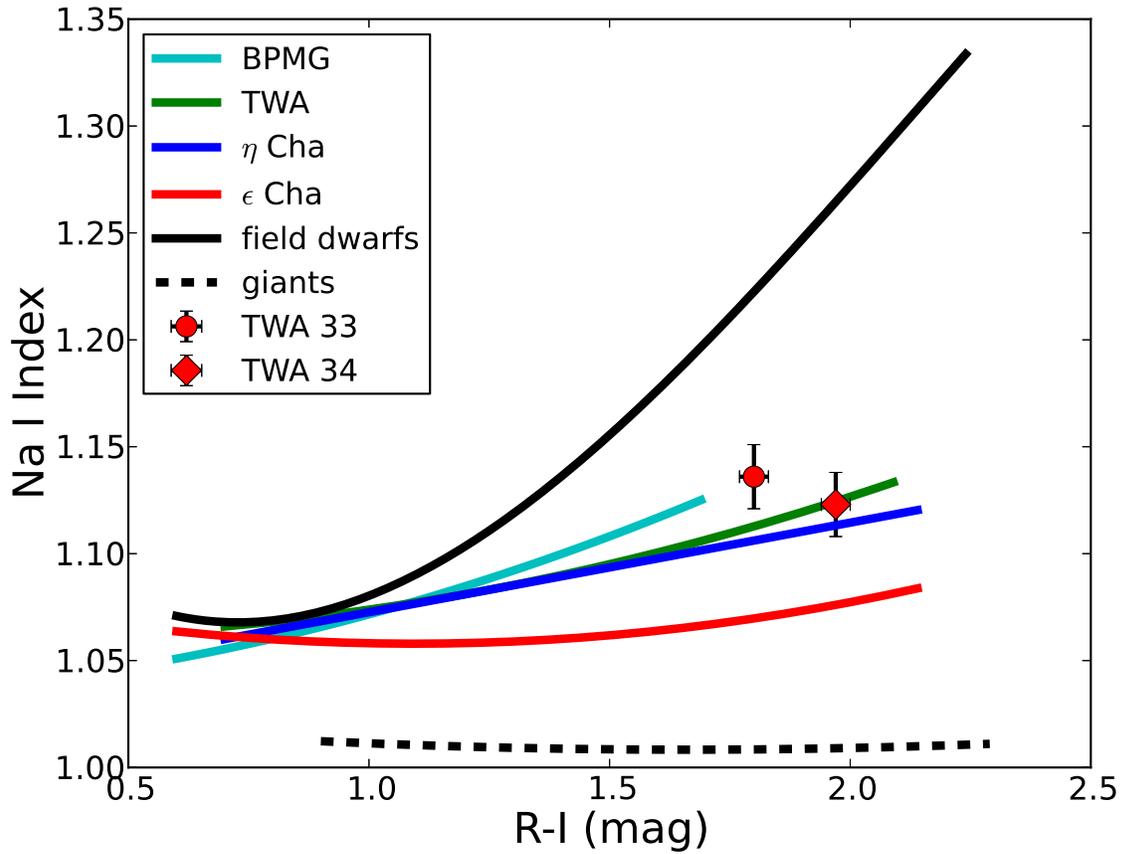}
\caption{Na I index trends for various stellar groups from \cite{law09}.  The location of TWA 33 and TWA 34 suggests a low surface
gravity, consistent with other TWA members.  Ages of these groups are as follows: Beta Pictoris Moving Group
$\sim$12 Myr, TWA $\sim$8 Myr, $\eta$ Cha $\sim$5-8 Myr, and $\epsilon$ Cha $\sim$ 6-7 Myr.}
\end{figure}
\begin{figure}
\plotone{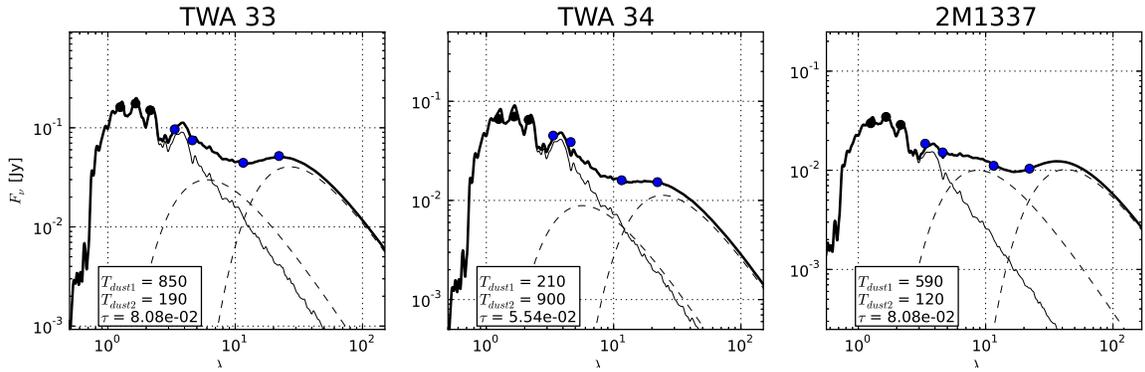}
\caption{Spectral energy distributions of TWA 33, TWA 34, and 2M1337 showing their mid-IR excess emission.  The solid
circles are photometric measurements from 2MASS and WISE. 
The thin black curve is an atmospheric model \citep{hau99} fit to the 2MASS data. The dashed lines are
single temperature blackbody dust fits to the excess emission with the dust temperatures (in K) 
as specified in the plot.  The thick curve is the spectral model +  blackbody dust fit. $\tau$ = infrared luminosity = $L_{IR}$/$L_{bol}$.}
\end{figure}
\clearpage

\end{document}